\documentclass[RNAAS]{aastex62}

\begin{document}

\newcommand{\kms}{\ensuremath{\mathrm{km}\,\mathrm{s}^{-1}}}
\newcommand{\galunits}{\ensuremath{\mathrm{km}\,\mathrm{s}^{-1}\,\mathrm{kpc}^{-1}}}
\newcommand{\galacc}{\ensuremath{\mathrm{km}^2\,\mathrm{s}^{-2}\,\mathrm{kpc}^{-1}}}
\newcommand{\MLsun}{\ensuremath{\mathrm{M}_{\sun}/\mathrm{L}_{\sun}}}
\newcommand{\Lsun}{\ensuremath{\mathrm{L}_{\sun}}}
\newcommand{\Msun}{\ensuremath{\mathrm{M}_{\sun}}}
\newcommand{\Ha}{\ensuremath{\mathrm{H}\alpha}}
\newcommand{\SFR}{\ensuremath{\mathit{SFR}}}
\newcommand{\aveSFR}{\ensuremath{\langle \mathit{SFR} \rangle}}
\newcommand{\sfrate}{\ensuremath{\mathrm{M}_{\sun}\,\mathrm{yr}^{-1}}}
\newcommand{\Aunits}{\ensuremath{\mathrm{M}_{\sun}\,\mathrm{km}^{-4}\,\mathrm{s}^{4}}}
\newcommand{\surfdens}{\ensuremath{\mathrm{M}_{\sun}\,\mathrm{pc}^{-2}}}
\newcommand{\voldens}{\ensuremath{\mathrm{M}_{\sun}\,\mathrm{pc}^{-3}}}
\newcommand{\gevcc}{\ensuremath{\mathrm{GeV}\,\mathrm{cm}^{-3}}}
\newcommand{\etal}{et al.}
\newcommand{\LCDM}{$\Lambda$CDM}
\newcommand{\ML}{\ensuremath{\Upsilon_*}}
\newcommand{\Mst}{\ensuremath{M_*}}
\newcommand{\Mg}{\ensuremath{M_g}}
\newcommand{\Mb}{\ensuremath{M_b}}
\newcommand{\gobs}{\ensuremath{\mathrm{g}_{\mathrm{obs}}}}
\newcommand{\gtot}{\ensuremath{\mathrm{g}_{\mathrm{tot}}}}
\newcommand{\gbar}{\ensuremath{\mathrm{g}_{\mathrm{bar}}}}
\newcommand{\azero}{\ensuremath{\mathrm{g}_{\dagger}}}

\title{A Precise Milky Way Rotation Curve Model \\ for an Accurate Galactocentric Distance}

\author{Stacy S. McGaugh}
\affil{Department of Astronomy, Case Western Reserve University, Cleveland, OH 44106}


\section{Introduction}

Many problems in Galactic structure benefit from accurate knowledge of the distance to the Galactic center, $R_0$. 
This crucial value has recently been measured to unprecedented accuracy
 --- $R_0 = 8.122 \pm 0.031\;\mathrm{kpc}$ --- 
thanks to relativistic effects observed during the pericenter passage of a star orbiting the central supermassive black hole \citep{GRAVITY}.
Combined with the observed proper motion of Sgr A$^*$ \citep{SgrAstar}, this Galactocentric distance implies a transverse speed of the sun of 
$245.58 \pm 1.32\;\kms$. Adopting $V_{\sun} = 12.24 \pm 0.47\;\kms$ \citep{solarmotion} for the solar motion  
leads to a circular speed of $\Theta_0 = 233.34 \pm 1.40\;\kms$ for the Local Standard of Rest.

In light of this development, I provide here an update to the Milky Way model of \citet{M16}.
In addition to the new Galactocentric distance, new measurements of the terminal velocities in the first quadrant are also available \citep{MGDQ1}.
Combining these with fourth quadrant terminal velocities \citep{MGDQ4} and the Galactic constants ($R_0$, $\Theta_0$) = (8.122 kpc, 233.3 \kms)
provides a remarkably detailed picture of the Galactic rotation curve.

The method of \citet{M16} applies the Radial Acceleration Relation \citep[RAR:][]{RAR} to 
derive the azimuthally averaged baryonic surface density profile $\Sigma(R)$ from the Galactic rotation curve. 
Features in $V(R)$ have corresponding features in $\Sigma(R)$ that map well to independently known features like the Centaurus spiral arm.
The resulting $\Sigma(R)$ departs from a pure exponential profile in a way that is typical of other spiral galaxies.

The model Q4MB of \citet{M16} provides the starting point here as its bulge model provides an excellent match to the inner rotation curve
obtained from the detailed modeling of \citet{Portail2015} when both are scaled to the same $R_0$. 
The pattern of bumps and wiggles at larger radii --- presumably the signature of massive spiral arms like
Centaurus --- is a good match to the terminal velocity data in both quadrants. The new, slightly larger $R_0$
makes the Milky Way more massive, a net increase of 9\% over the Q4MB model, with correspondingly higher surface densities.

\section{Results}

\begin{figure}
\plotone{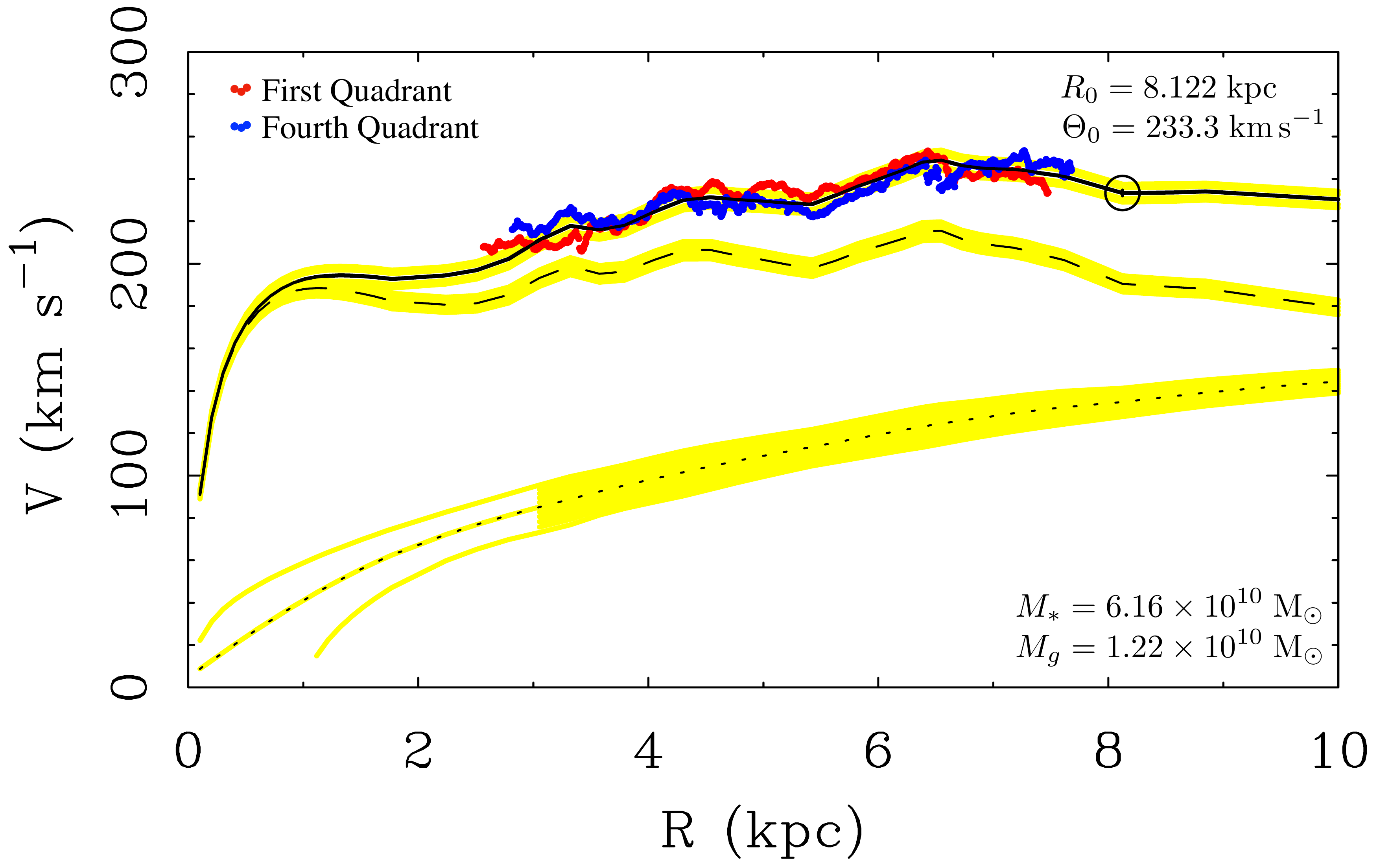}
\caption{Model rotation curve for the Milky Way (solid line) matched to the first \citep[red points;][]{MGDQ1} and 
fourth \citep[blue points;][]{MGDQ4} quadrant terminal velocities following \citet{M16}. 
The dashed line shows the rotation curve of the baryonic component and the dotted line is the implied dark matter.
Yellow bands illustrate the range of variation for $\pm$5\% in stellar mass. 
}
\label{fig:MW}
\end{figure}

The updated model is shown in Fig.\ \ref{fig:MW}.
In addition to fitting the terminal velocities, it also fits the highly accurate $\Theta_0$.
This implies a slight dip in the rotation curve \citep{SofueDip}, as expected from the sun's location interior to the Perseus arm.
The rotation curve outside the solar circle is not constrained by the terminal velocities, but the model extrapolates well
to known constraints \citep{M16}, including that from the recent analysis of the GD-1 stellar stream \citep{MalhanIbata}.

The total stellar and gas mass of the model is $\Mst = 6.16 \times 10^{10}\;\Msun$ and $\Mg = 1.22 \times 10^{10}\;\Msun$. 
Treating the asymmetry between quadrants as an uncertainty, \Mst\ is known to $\sim$5\%.
This is a map of the gravitational potential of each mass component, including the bumps and wiggles due to spiral structure in the disk. 

The Milky Way is strongly baryon dominated interior to the sun, satisfying any definition of maximum disk \citep{Sackett,StarkmanMaxDisk}.
The implied dark matter halo is also well constrained. 
The numerical results can be approximated by a pseudo-isothermal halo 
with core radius $R_C = 3.05$ kpc and asymptotic velocity $V_{\infty} = 185.8\;\kms$. 
The nearest NFW approximation has $c_{200} = 8.07$ and $V_{200} = 161.5\;\kms$.

The local density of dark matter is 
$\rho_{DM}(R_0) = 6.76^{+0.08}_{-0.14} \times 10^{-3}\; \voldens = 0.257^{+0.003}_{-0.005}\; \gevcc$.
The stated uncertainty is for a 5\% uncertainty in \Mst. 
This estimate from the \textit{radial} force is about half that found by studies of the 
\textit{vertical} force \citep[e.g.,][]{Beinayme}.
This may be a hint about the flattening of the halo or other effects \citep[e.g.,][]{DMequil}. 

It will be interesting to compare these results with Gaia, as the information utilized here is entirely independent.


\end{document}